\documentclass[prb,amsmath,amssymb,showpacs]{revtex4}
\usepackage{xspace}

\usepackage{graphics}
\usepackage{graphicx}

\newcommand{\tlabel}[1]{}
\newcommand{\elabel}[1]{\label{eq:#1}}
\newcommand{\eref}[1]{(\ref{eq:#1})}
\newcommand{\Eref}[1]{Eq.~\eref{#1}}

\newcommand{\Fref}[1]{Fig.~\ref{fig:#1}}
\newcommand{\flabel}[1]{\label{fig:#1}}

\newcommand{\slabel}[1]{\label{sec:#1}}

\newcommand{\Sref}[1]{Section~\ref{sec:#1}}

\newcommand{\Subsref}[1]{Subsection~\ref{sec:#1}}

\newcommand{\ie}{{\it i.e.}\xspace}

\newcommand{\etal}{{\it et}~{\it al.}\xspace}

\newcommand{\onebody}{}

\renewcommand{\emph}[1]{{\it #1}}

\newcommand{\gvec}[1]{\mathbf{#1}}
\newcommand{\zerovec}{\gvec{0}}

\newcommand{\kvec}{\gvec{k}}
\newcommand{\rvec}{\gvec{r}}
\newcommand{\nvec}{\gvec{n}}
\newcommand{\mvec}{\gvec{m}}
\newcommand{\jvec}{\gvec{j}}

\newcommand{\dint}[1]{\mathchoice{\!\!\mathrm{d}#1\,}{\!\mathrm{d}#1\,}{\!\mathrm{d}#1\,}{\!\mathrm{d}#1\,}}
\newcommand{\ddint}[1]{\mathchoice{\!\!\mathrm{d}^d#1\,}{\!\mathrm{d}^d#1\,}{\!\mathrm{d}^d#1\,}{\!\mathrm{d}^d#1\,}}

\newcommand{\rtilde}{\widetilde{\rvec}}
\newcommand{\rhat}{\widehat{\rvec}}
\newcommand{\rtildeInt}{\widetilde{r}}
\newcommand{\xtilde}{\widetilde{x}}

\newcommand{\imag}{\imath}

\newcommand{\FC}{\mathcal{F}}
\newcommand{\WC}{\mathcal{W}}
\newcommand{\ZC}{\mathcal{Z}}
\newcommand{\PS}{\ZC}
\newcommand{\Fpot}{\FC}
\newcommand{\Wpot}{\WC}
\newcommand{\WpotNoneq}{\widetilde{\WC}}

\newcommand{\ideal}{\text{\tiny id.}}
\newcommand{\Fpotid}{\FC_{\ideal}}

\newcommand{\uid}{u_{\ideal}}

\newcommand{\ave}[1]{\left\langle #1 \right\rangle\!}

\newcommand{\half}{\frac{1}{2}}

\newcommand{\fourth}{\frac{1}{4}}
\newcommand{\quarter}{\fourth}
\newcommand{\Rset}{\mathbb{R}}
\newcommand{\Sset}{\mathbb{S}}
\newcommand{\Zset}{\mathbb{Z}}
\newcommand{\Kset}{\mathbb{K}}
\newcommand{\Iset}{\mathbb{I}}
\newcommand{\ID}{\mathbf{1}}

\newcommand{\bra}[1]{\left\langle#1\right|}
\newcommand{\ket}[1]{\left|#1\right\rangle}
\newcommand{\Tr}{\operatorname{Tr}}

\begin{document}
\title{Phase field model of interfaces in single-component systems
derived from classical density functional theory}
\author{Gunnar Pruessner}
\affiliation{
 University of Warwick,
 Mathematics Institute,
 Gibbet Hill Road,
 Coventry CV4 7AL, UK}
\affiliation{
 Imperial College London,
 Department of Physics,
 Exhibition Road,
 London SW7 2AZ, UK}
\author{A. P. Sutton}
\affiliation{
 Imperial College London,
 Department of Physics,
 Exhibition Road,
 London SW7 2AZ, UK}
\date{\today}

\begin{abstract}
Phase field models have been applied in recent years to grain
boundaries in single-component systems. The models are based on the
minimization of a free energy functional, which is constructed
phenomenologically rather than being derived from first principles.
In single-component systems the free energy is a functional of a
``phase field'', which is an order parameter often referred to as
the crystallinity in the context of grain boundaries, but with no
precise definition as to what that term means physically.
We present a derivation of the phase field model by Allen and Cahn
from classical density functional theory first for crystal-liquid interfaces
and then for grain boundaries. The derivation provides a clear
physical interpretation of the phase field, and it sheds light on
the parameters and the underlying approximations and limitations of
the theory. We suggest how phase field models may be improved.
\end{abstract}

\maketitle

\section{Introduction}

The classical density functional theory (DFT) advanced by Haymet and Oxtoby
\cite{HaymetOxtoby:1981,OxtobyHaymet:1982} was developed to describe
crystal-liquid interfaces. In this paper we will show how this theory
may be approximated in a transparent way to derive phase field models
of crystal-liquid and grain boundary interfaces in single-component
systems. We will also discuss how classical
DFT may be applied to grain
boundaries in single-component systems. At specified temperature and chemical
potential, atoms at the boundary are free to rearrange themselves to
accommodate a fixed change of crystal orientation on either side of
the boundary. In principle DFT provides an exact treatment of these
defects at the atomic level within a grand canonical ensemble.
The application of DFT to grain boundaries is of
considerable interest in itself as an alternative to grand canonical
Monte Carlo simulations and molecular dynamics simulations.

Grain boundaries in single-component systems have been modeled
using phase field approaches, most recently by Carter and coworkers
\cite{KobayashiWarrenCarter:2000,TangCarterCannon:2006} and Nestler
and Wheeler \cite{NestlerWheeler:2002}. All of these studies make
use of an Allen-Cahn type free energy functional, obtained in the
case of Carter et al. after first integrating out an ``orientation
field''. These models are based on the minimization of a free
energy, which is constructed phenomenologically, referring to
symmetries, relevant physical parameters, effective interactions
etc., rather than being derived from first principles. The free
energy is a functional of a ``phase field'', which is an order
parameter often referred to as the ``crystallinity'' in the context
of grain boundaries. The crystallinity is interpreted as a measure
of the local degree of structural order, but with no attempt to
define it more precisely.

Our aim in this paper is to derive a phase field model for crystal-liquid interfaces and grain
boundaries in single-component systems from classical density
functional theory, providing not only a clear physical
interpretation of the phase field, but also an exposition of the
various approximations along the way. The mapping also enables us to
identify limitations of phase field modeling of interfaces, and to suggest how it may be improved.

We begin by outlining classical DFT and its application by Haymet
and Oxtoby to crystal-liquid interfaces, noting key assumptions. While
this derivation focuses on a thermodynamic self-consistency
equation, applying the same procedure directly to the grand
potential leads to the Allen-Cahn equation and subsequently to a
phase field model for grain boundaries recently introduced by Carter
\etal. \cite{KobayashiWarrenCarter:2000}. An interesting by-product of
the derivation of a phase field model for grain boundaries is the result
that the energy and width of a boundary are both predicted to be reduced
when there is a short common vector in the reciprocal lattices of both
crystals.

\section{Classical density functional theory}
Following an integration over the momentum degrees of freedom, the
canonical partition sum\cite{HansenMcDonald:2006} in $d$ dimensions
for $N$ identical classical particles at positions
$\rvec_1,\ldots,\rvec_N\in\Omega$, where $\Omega$ is an arbitrarily
large volume,
interacting via a potential
$V_N(\rvec_1,\ldots,\rvec_{\nvec})$ in an external potential
$U(\rvec)$ is
\begin{equation}
Z_N = \frac{1}{N!} \Lambda^{-dN} \int_\Omega \!\! d^d\mathrm{r}_1\ldots
d^d\mathrm{r}_N\ \exp\left(-\beta \big(V_N(\rvec_1,\ldots,\rvec_{\nvec}) +
\sum_{i=1}^N U(\rvec_i)\big)\right),
\end{equation}
where
\begin{equation}
\Lambda=\left(\frac{\beta h^2}{2 m \pi}\right)^{1/2}
\elabel{deBroglie_length}
\end{equation}
is the de Broglie thermal wave length. The Laplace transform of the
canonical partition sum is the grand canonical partition sum
\begin{equation}
\PS = \sum_{N=0}^\infty e^{\beta \mu N} Z_N \ .
\end{equation}
This suggests the introduction of the dimensionless one-body local
potential, not to be confused with the effective one-particle
potential introduced later:
\begin{equation}
u(\rvec)=\beta \mu - \beta U(\rvec).
\end{equation}
The dimensionless grand potential, $\Wpot=\Wpot[u] = - \ln \PS$, is
then a functional of the local potential $u$ only, apart from the
implicit dependence on the interaction potential, and the
temperature which is assumed to be constant.

Introducing an operator $\hat{\rho}_N(\rvec;
\rvec_1,\ldots,\rvec_{\nvec})=\sum_i \delta(\rvec-\rvec_i)$ reveals
that the particle density, which is the ensemble average
$\ave{\cdot}$ of this operator, is given by functional
differentiation of the grand potential with respect to the local
potential
\begin{equation}
\rho([u]; \rvec) = \ave{\hat{\rho}_N(\rvec;
\rvec_1,\ldots,\rvec_{\nvec})} = - \frac{\delta \Wpot[u]}{\delta
u(\rvec)}.
\end{equation}
This can be seen by rewriting $\beta \mu N -\beta
\sum_{i=1}^N U(\rvec_i)$ as
$\int_\Omega \ddint{r} u(\rvec) \hat{\rho}_N(\rvec; \ldots)$
in the grand canonical partition sum.

Taking a Legendre transform, the free energy
\begin{equation}
\Fpot[\rho] = \Wpot[u] + \int_\Omega \ddint{r'} \rho(\rvec') u(\rvec')
\elabel{F_from_W}
\end{equation}
produces the conjugates of the cumulants of the density, in particular
\begin{equation}
u(\rvec) = u([\rho]; \rvec) = \frac{\delta}{\delta \rho(\rvec)}
\Fpot[\rho].
\end{equation}

The free energy of an ideal, \ie non-interacting system, where
$V_N\equiv0$, can be
integrated,
\begin{equation}
\Fpotid[\rho] =
\int_\Omega\ddint{r}\rho(\rvec)
\left( \ln(\Lambda^d \rho(\rvec)) - 1 \right)
\elabel{def_Fpotid}
\end{equation}
and immediately gives rise to the barometric formula
$\rho(\rvec)_{\ideal} = \Lambda^{-d} \exp(\uid(\rvec))$. In an ideal
system, the external potential necessary to produce a given density
profile is found simply by inverting the barometric formula. For a
given density profile $\rho$ observed in an interacting particle
system, one can calculate the \onebody local potential that would
be needed in an ideal system, $\uid$, to produce the same density
profile. The difference between the \onebody local potential, $u$,
operating in the interacting system, and the ideal \onebody local
potential is the effective one particle potential $C([\rho];
\rvec)$, defined as
\begin{equation}
C([\rho]; \rvec) = \uid([\rho]; \rvec) - u([\rho]; \rvec) =
\ln(\Lambda^d \rho(\rvec)) - u([\rho]; \rvec). \elabel{C_as_diff}
\end{equation}
In other words, an ideal system with local potential $u+C$ has the same
density profile $\rho$ as the interacting system with local potential
$u$.

Similarly, one defines the excess free energy
\begin{equation}
-\Phi[\rho] = \Fpot[\rho]-\Fpotid[\rho] \elabel{def_Phi},
\end{equation}
and finds by differentiation
\begin{equation}
\frac{\delta}{\delta \rho(\rvec)} \Phi[\rho] = C([\rho]; \rvec).
\elabel{def_eopp}
\end{equation}
Further differentiation of the effective one-particle potential produces
higher order correlation functions. In particular, the direct
correlation function $C^{(2)}([\rho]; \rvec, \rvec')$ is defined as
\begin{equation}
C^{(2)}([\rho]; \rvec, \rvec') \equiv
\frac{\delta^2 }{\delta \rho(\rvec)\delta \rho(\rvec')}
\Phi[\rho] \ .
\elabel{def_C2}
\end{equation}
\Eref{def_C2} immediately suggests an expansion of the excess free
energy about a reference state, which in the following will be the
\emph{infinite homogeneous liquid} with constant density $\rho_0$,
\begin{equation}
\Phi[\rho] = \Phi[\rho\equiv\rho_0]
+ \int_\Omega\ddint{r}C([\rho\equiv\rho_0]; \rvec)
(\rho(\rvec)-\rho_0)
+ \half \int_\Omega\ddint{r}\; \ddint{r'} C^{(2)}([\rho\equiv\rho_0]; \rvec, \rvec')
(\rho(\rvec)-\rho_0)
(\rho(\rvec')-\rho_0)
+ \ldots
\elabel{expansion_Phi}
\end{equation}
We use this expansion below to derive features of the crystal phase
from the reference system.\footnote{Note that an expansion about a
crystal reference system is in principle possible, but breaks
symmetry: given a direct correlation function that necessarily lacks
$O(3)$ symmetry for the crystalline state, one cannot expect
\Eref{expansion_Phi} to be invariant under $O(3)$ if it is
truncated. However, in the absence of an anisotropic external
potential, the excess free energy must remain unchanged under
rotation of the density $\rho$ in space. Therefore, in general, one
cannot use a crystal reference state.}

Using the Ornstein-Zernike equation \cite{HansenMcDonald:2006} the
Fourier transform $c(\kvec)$ of the two-point direct correlation
function over a domain with volume $V''$ in a translationally invariant
system, see \Eref{def_ck}, can be directly related to the structure
factor $S(\kvec)$
\begin{equation}
S(\kvec) = \left(1-\rho_0 V'' c(\kvec)\right)^{-1} \ ,
\elabel{structure_factor}
\end{equation}
so that the direct correlation function has a direct physical and
experimental meaning. It is, in fact, the \emph{only} link to
experiment, which raises the question of how it is possible that a
perturbation theory about a \emph{liquid} can describe features of a
\emph{crystal}. For example, one might doubt that the tetrahedral
structure of crystalline silicon could be predicted from the two point
direct correlation \emph{function} of liquid silicon, which has an
average coordination number of $6$. On the other hand, the
\emph{functional} $C^{(2)}([\rho]; \rvec, \rvec')$, \ie the direct
correlation function including its functional dependence on the
density within the \emph{entire} system, contains all information
about higher order correlations through functional differentiation.
In \Eref{expansion_Phi} $C^{(2)}$ is evaluated only at
$\rho(\rvec)\equiv\rho_0$, so the full functional dependence is
suppressed. A key assumption of the theory is that by a judicious
choice of $\rho_0$ the expansion in \Eref{expansion_Phi} can be made
acceptable. This debate has appeared prominently in the literature
\cite{RamakrishnanYussouff:1979}. On a
more technical level, the functional Taylor series
\Eref{expansion_Phi} may have a finite ``radius of convergence'' in
$\rho(\rvec)-\rho_0$, not extending beyond the liquid phase, not
least because a phase transition introduces singularities.

\section{Application to interfaces}
\slabel{App_to_Ints}
In all that follows the approximations are based on the two point
direct correlation function $C^{(2)}([\rho\equiv\rho_0]; \rvec,
\rvec')$, which is a function only of the separation $|\rvec-\rvec'|$
in a homogeneous system. To ease notation, the homogeneous, infinite
reference liquid will be denoted by subscript $0$, instead of
explicitly carrying the argument $\rho\equiv\rho_0$. In this
notation, the one-particle potential of the reference system, which
is assumed to be homogeneous and
to have vanishing \onebody local potential
$u(\rvec)\equiv u_0=\beta\mu$, is related to its density by
\begin{equation}
C_0 = \ln(\Lambda^d \rho_0) - \beta \mu
\elabel{self_consistency_reference}
\end{equation}
consistent with \Eref{C_as_diff}.

We use the reference system to calculate features of an interfacial
system (distinguished by subscript $i$) for a given \onebody local
potential $u_i(\rvec)$, which amounts to finding a root $\rho_i(\rvec)$
of $u_i=\delta \Fpot_i[\rho]/\delta \rho$. Equivalently one can
minimize the functional
\begin{equation}
\WpotNoneq([\rho],[u_i])=\Fpot_i[\rho]-\int_\Omega \ddint{r} u_i(\rvec)
\rho(\rvec)
\elabel{def_WpotNoneq}
\end{equation}
with respect to $\rho$. At the minimum $\rho\equiv\rho_i$ and
$\WpotNoneq([\rho_i],[u_i])$ reduces to the grand potential, as seen
in \Eref{F_from_W}. For a vanishing \onebody local potential in the
interfacial system $u_i\equiv \beta\mu$. The expansion in
\Eref{expansion_Phi} together with
\Eref{self_consistency_reference}, \Eref{def_Fpotid} and
\Eref{def_Phi}, is the starting point for the DFT used by Haymet and
Oxtoby \cite{HaymetOxtoby:1981}:
\begin{eqnarray}\elabel{Wpoti_DFT}
\WpotNoneq([\rho_i],[u_i\equiv\beta\mu]) &=&
\int_\Omega \ddint{r'} \left[ \ln \left(\rho_i(\rvec')/\rho_0\right) - 1 \right]
\rho_i(\rvec') - \Phi_0  + \int_\Omega \ddint{r'} C_0 \rho_0\\
&&-\half \int_\Omega \ddint{r'}\!\! \int_\Omega\ddint{r''}
C^{(2)}_0(|\rvec''-\rvec'|) \left(\rho_i(\rvec')-\rho_0\right)
\left(\rho_i(\rvec'')-\rho_0\right). \nonumber
\elabel{2ndorder}
\end{eqnarray}
It is worth stressing that the only approximation made so far is the
truncation of the functional Taylor series \Eref{expansion_Phi} at
second order.

To derive the results of
[\onlinecite{HaymetOxtoby:1981,OxtobyHaymet:1982}]
one requires that $\WpotNoneq$ differentiated with respect to
$\rho_i$ vanishes. Using $C^{(2)}_0(\rvec)=C^{(2)}_0(-\rvec)=C^{(2)}_0(|\rvec|)$ this
produces the self-consistency equation
\begin{equation}
0 = \ln \rho_i(\rvec)/\rho_0 - \int_\Omega \ddint{r'}
C^{(2)}_0(|\rvec -\rvec'|) \left(\rho_i(\rvec')-\rho_0\right),
\elabel{self_consistency}
\end{equation}
which can also be obtained by expanding the effective one particle
potential in a functional Taylor series. It is very difficult to
find a root $\rho_i$ for this equation satisfying particular
boundary conditions far from the interface.

Physically, it is very appealing to represent the density in a
pseudo-Fourier sum \cite{HaymetOxtoby:1981,RamakrishnanYussouff:1979}
\begin{equation}
\rho_i(\rvec)=\rho_0 \left(
1 + \sum_{\nvec} \mu_{\nvec}(\rvec) e^{\imag \kvec_{\nvec} \rvec}
\right)
\elabel{reparam_rho}
\end{equation}
with reciprocal lattice vectors $\kvec_\nvec$ indexed by
$\nvec\in\Zset^d$. For simplicity we have assumed the crystal has a monatomic basis.
In crystals with more than one atom in the basis there will be
additional phase factors in the pseudo-Fourier expansion. At a crystal-liquid interface there is only
one crystal lattice, but at a grain boundary there is one set of reciprocal
lattice vectors for each crystal, and the expansion in
\Eref{reparam_rho} is over the union of reciprocal lattice
vectors of the two crystals. In the following
derivation we will not make use of any specific properties of
$\{\kvec_\nvec\}$ other than its discreteness and completeness (see
\Eref{self_consistency_family}), as well as its orthogonality (see
\Eref{self_consistency_expanded_FT}), which all necessitate a finite
Fourier domain, introduced in \Eref{def_scalar_product} as the
volume $V$. For grain boundaries it follows that the derivation
applies only to cases where a three-dimensional
coincidence site lattice (CSL) \cite{SuttonBalluffi:1995} exists with a reasonably small unit cell.

Owing to the spatial dependence of the coefficients
$\mu_{\nvec}(\rvec)$, \Eref{reparam_rho} is not a true Fourier sum.
However, it can accommodate every density $\rho_i(\rvec)$, even if
the density is not periodic. While the spatial dependence has, \emph{a
priori}, the
unfortunate consequence that individual coefficients $\mu_{\nvec}(\rvec)$
cannot be calculated by projecting $\rho_i(\rvec)$ onto the functions
$e^{-\imag \kvec_{\nvec} \rvec}$, the spatial dependence of the
coefficients can be thought of as a means to \emph{separate length
scales}. The atomic-scale density variations are captured by the
exponential $\exp(\imag \kvec_\nvec \rvec)$, while more gradual
changes in density are captured by $\mu_{\nvec}(\rvec)$. We discuss
below how the $\mu_{\nvec}(\rtilde)$ for different $\rtilde$ can be
obtained by taking a local Fourier integral within a volume
$V(\rtilde)$,
\begin{equation}
\mu_{\nvec}(\rtilde) \rho_0=V^{-1} \int_{V(\rtilde)} \!\!\! \ddint{r} \rho_i(\rvec)
e^{- \imag \kvec_{\nvec} \rvec} \quad\text{ for }\quad n\ne0
\elabel{def_scalar_product}
\end{equation}
and $\rho_0 (1+\mu_{\zerovec}(\rtilde))=V^{-1} \int_{V(\rtilde)} \ddint{r} \rho_i(\rvec)$.
Here $V(\rtilde)$ denotes the
domain the integral is running over, a
parallelepiped of volume $V$ which is a unit cell of the
coincidence site lattice or a multiple thereof.

Using the new representation \Eref{reparam_rho} in the
self-consistency equation \eref{self_consistency} and expanding the
$\mu_\nvec(\rvec')$ about $\rvec$ gives for \Eref{self_consistency}
\begin{equation}
\ln \left(1+\sum_\nvec \mu_\nvec(\rvec) e^{\imag \kvec_\nvec
\rvec}\right)
=
V''\rho_0 \sum_{\nvec} e^{\imag \kvec_\nvec \rvec} \left[
   \mu_\nvec(\rvec) c(\kvec_\nvec)
-  \imag (\nabla_\rvec \cdot \nabla_\kvec) \mu_\nvec(\rvec) c(\kvec_\nvec)
-  \half (\nabla_\rvec \cdot \nabla_\kvec)^2 \mu_\nvec(\rvec) c(\kvec_\nvec)
+ \ldots \right]
\elabel{self_consistency_expanded}
\end{equation}
where
\begin{equation}\elabel{cn_Omega}
\int_\Omega\ddint{r'} C^{(2)}_0(|\rvec-\rvec'|) e^{-\imag \kvec
(\rvec-\rvec')} =
V'' c(\kvec)
\end{equation}
has been used, based on the definition
\begin{equation}
c(\kvec)=V^{\prime\prime-1} \int_{V''}\!\! \ddint{r''} e^{-\imag \kvec
\rvec''} C^{(2)}_0(\rvec'')\ .
\elabel{def_ck}
\end{equation}
The central idea is that $V''$ is so large that $C^{(2)}$ vanishes
outside this volume and that $\Omega$ is sufficiently large that the
failure of \Eref{cn_Omega} due to the shift by $\rvec$ when writing
\Eref{cn_Omega} as \Eref{def_ck} is negligible.

So far, only two approximations have been made, the functional
Taylor series of the excess free energy and the Taylor series for
the pseudo-Fourier coefficients $\mu_\nvec$. It is perhaps not
surprising that \Eref{self_consistency_expanded} remains just as
intractable as \Eref{self_consistency}, which becomes obvious by
taking a Fourier transform over the domain $V(\rtilde)$ centered at
$\rtilde\in\Omega$ on both sides of
\Eref{self_consistency_expanded}:
\begin{multline}
V^{-1} \int_{V(\rtilde)}\!\!\!\ddint{r} e^{-\imag \kvec_\mvec \rvec}
\ln \left(1+\sum_\nvec \mu_\nvec(\rvec) e^{\imag \kvec_\nvec
\rvec}\right)
=\\
V^{-1} \int_{V(\rtilde)}\!\!\!\ddint{r} e^{-\imag \kvec_\mvec \rvec}
V''\rho_0 \sum_{\nvec} e^{\imag \kvec_\nvec \rvec} \left[
   \mu_\nvec(\rvec) c(\kvec_\nvec)
-  \imag (\nabla_\rvec \cdot \nabla_\kvec) \mu_\nvec(\rvec)
c(\kvec_\nvec) -  \half (\nabla_\rvec \cdot \nabla_\kvec)^2
\mu_\nvec(\rvec) c(\kvec_\nvec) + \ldots \right].
\elabel{self_consistency_expanded_FT}
\end{multline}
The problem is the $\rvec$-dependence of the coefficients
$\mu_\nvec(\rvec)$ which renders the orthogonality of the
exponentials on the right hand side unexploitable. On the other
hand, the representation in \Eref{reparam_rho} has not yet been
fully exploited, in particular, no use has been made so far of the
many degrees of freedom in the parametrization.

\begin{figure}
\begin{center}
\includegraphics*[width=0.6\linewidth]{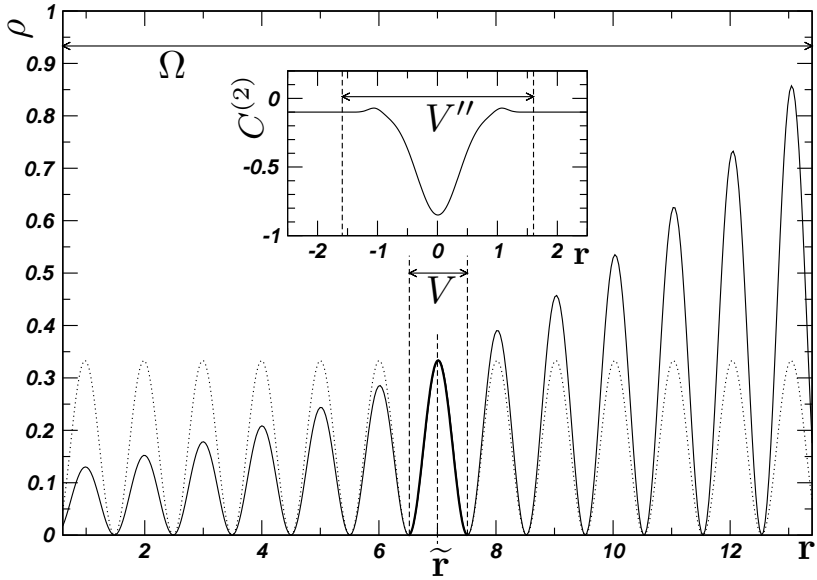}
\end{center}
\caption{\flabel{local_Fourier} Cartoon of a local Fourier
transform. The full line shows the actual density profile
$\rho(\rvec)$ within a system of size $\Omega$ parameterized by an
expression like \Eref{reparam_rho} which contains a space dependent
pseudo-Fourier coefficient. A Fourier transform taken over the small
region $V$ marked by the dashed lines around $\rtilde$, produces
coefficients that reproduce the density locally (shown as thick
line) and can be continued periodically throughout the system
(dotted line). The inset shows the two point direct correlation
function $C^{(2)}(\rvec)$ which drops to a negligible value outside
the domain $V''$.}
\end{figure}

Haymet and Oxtoby achieved a separation of length scales by
utilizing the parametrization \Eref{reparam_rho}. Firstly, they
introduced a family of density profiles indexed by $\rtilde$,
\begin{equation}
\rho(\rvec;\rtilde) = \rho_0
\left(
1 + \sum_{\nvec} \mu_{\nvec}(\rtilde) e^{\imag \kvec_{\nvec} \rvec}
\right)\ ,
\elabel{reparam_rho_u}
\end{equation}
which coincides with \Eref{reparam_rho} for $\rtilde=\rvec$.
Secondly, they replace $\mu_\nvec(\rvec)$ by $\mu_\nvec(\rtilde)$ in
\Eref{self_consistency_expanded}, requiring that the
$\mu_\nvec(\rtilde)$ at fixed $\rtilde$ are solutions of a slightly
different problem,
\begin{equation}
\ln \left(1+\sum_\nvec \mu_\nvec(\rtilde) e^{\imag \kvec_\nvec
\rvec}\right)
=
V''\rho_0 \sum_{\nvec} e^{\imag \kvec_\nvec \rvec} \left[
   \mu_\nvec(\rtilde) c(\kvec_\nvec)
-  \imag (\nabla_\rvec \cdot \nabla_\kvec) \mu_\nvec(\rtilde)
c(\kvec_\nvec) -  \half (\nabla_\rvec \cdot \nabla_\kvec)^2
\mu_\nvec(\rtilde) c(\kvec_\nvec) + \ldots \right],
\elabel{self_consistency_expanded_family}
\end{equation}
and therefore
\begin{multline}
V^{-1} \int_{V(\rtilde)}\!\!\!\ddint{r} e^{-\imag \kvec_\mvec \rvec}
\ln \left(1+\sum_\nvec \mu_\nvec(\rtilde) e^{\imag \kvec_\nvec
\rvec}\right)
=\\
V^{-1} \int_{V(\rtilde)}\!\!\!\ddint{r} e^{-\imag \kvec_\mvec \rvec}
V''\rho_0 \sum_{\nvec} e^{\imag \kvec_\nvec \rvec}
\left[
   \mu_\nvec(\rtilde) c(\kvec_\nvec)
-  \imag (\nabla_\rvec \cdot \nabla_\kvec) \mu_\nvec(\rtilde) c(\kvec_\nvec)
-  \half (\nabla_\rvec \cdot \nabla_\kvec)^2 \mu_\nvec(\rtilde) c(\kvec_\nvec)
+ \ldots \right] \\
=V''\rho_0
\left[
   \mu_\mvec(\rtilde) c(\kvec_\mvec)
-  \imag (\nabla_\rvec \cdot \nabla_\kvec) \mu_\mvec(\rtilde)
c(\kvec_\mvec) -  \half (\nabla_\rvec \cdot \nabla_\kvec)^2
\mu_\mvec(\rtilde) c(\kvec_\mvec) + \ldots \right],
\elabel{self_consistency_family}
\end{multline}
corresponding to \Eref{self_consistency_expanded_FT}. For a
complete, discrete set of reciprocal lattice vectors
$\{\kvec_\nvec\}$ \Eref{self_consistency_family} is equivalent to
\Eref{self_consistency_expanded_family}; in other words,
\Eref{self_consistency_expanded_family} is implied by
\Eref{self_consistency_family} only if \eref{self_consistency_family}
applies to the entire set of reciprocal lattice vectors.

The solutions of \Eref{self_consistency_family} for each fixed
$\rtilde$ are periodic density profiles (see \Fref{local_Fourier})
of systems in a certain periodic external potential and it is
plausible to assume that solutions in the form $\mu_\nvec(\rtilde)$
exist. Moreover, \Eref{self_consistency_expanded_family} evaluated
at $\rtilde=\rvec$ coincides with \Eref{self_consistency_expanded},
so that any set of solutions of \Eref{self_consistency_family} also
represents a solution of \Eref{self_consistency_expanded_FT}. In
other words, Haymet and Oxtoby \emph{construct} a family of physical
systems parameterized by $\rtilde$ in an external potential, the
density profiles of which are given by $\mu_\nvec(\rtilde)$ and can
be used to construct a solution of the interfacial problem
\Eref{self_consistency_expanded_FT} by evaluating
\Eref{reparam_rho_u} at $\rtilde=\rvec$.

Because \Eref{self_consistency_family} can also be obtained by
assuming in \Eref{self_consistency_expanded_FT} the
$\mu_\nvec(\rvec)$ vary so little within $V(\rtilde)$ they can be
replaced by $\mu_\nvec(\rtilde)$ (see \Eref{def_scalar_product}), going from
\Eref{self_consistency_expanded_FT} to
\Eref{self_consistency_family} is equivalent to a separation of
length scales.
This separation of length scales validates \Eref{def_scalar_product}.
It is illustrated in
\Fref{local_Fourier}, where the Fourier coefficient $\mu_\nvec(\rtilde)$
obtained by taking a Fourier integral over a small domain centered at
$\rtilde$, gives rise to a density profile that can be periodically
continued throughout the system, but coincides almost perfectly with the actual density
$\rho_i(\rvec)$ within $V(\rtilde)$.

Small $V$ are desirable so that 
the separation of length scales as encapsulated in
\Eref{def_scalar_product} applies
with no or only small corrections. For a grain boundary, the volume $V$ is
a unit cell of the CSL, so that large $\Sigma$
boundaries are expected to be less reliably treated. Similarly,
small $V''$ are desirable so that the expansion in
\Eref{self_consistency_expanded} has only small corrections, which
implies that shorter-ranged direct correlation functions are easier
to handle in this formalism. The size of both volumes is to be
compared to the scale on which $\mu_\nvec(\rvec)$ change.

We see in \Eref{self_consistency_family} that the separation of
length scales embodied in \Eref{reparam_rho_u} simplifies the
original equation \eref{self_consistency_expanded_FT} dramatically
by enabling the orthogonality of the functions $\exp(\imag
\kvec_\nvec \rvec)$
to be
exploited.
Instead of finding roots $\mu_\nvec(\rvec)$ for all $\kvec_\nvec$ and
all $\rvec\in\Omega$ of the original integral equation
\Eref{self_consistency_expanded_FT} simultaneously, the integro-differential equation
\Eref{self_consistency_family} can in principle be integrated, as both
sides are local in $\rtilde$. As shown in ref.~[\onlinecite{OxtobyHaymet:1982}],
the $\mu_\nvec(\rvec)$ can be considered positions of particles in a
(complicated) potential at ``time'' $\rvec$ projected on the interface
normal.
In DFT, the next steps would consist in determining the
necessary degrees of freedom, incorporating all symmetries and
prescribing a procedure to find the roots $\mu_\nvec(\rtilde)$ for every
$\nvec$ and $\rtilde$.

\section{Derivation of phase field models}

\slabel{deri_PFM}

\subsection{Crystal-liquid interfaces}
We will now build on the approximations, expansions and
parameterizations described in the previous section to derive an Allen-Cahn phase field model for
crystal-liquid interfaces from the grand-potential \Eref{Wpoti_DFT}. In the next sub-section we will
derive a phase field model for grain boundaries.

The grand potential $\WpotNoneq$ in the form \Eref{Wpoti_DFT} is
reparameterized using \Eref{reparam_rho} and the integrals over
$\Omega$ rewritten as $\int_\Omega \ddint{r'} =
\int_\Omega\ddint{\rtildeInt} V^{-1}\!\!\int_{V(\rtilde)}\ddint{r'}$:
\begin{eqnarray}
\WpotNoneq([\rho_i],[u_i\equiv\beta\mu]) &=&
\int_\Omega\ddint{\rtildeInt}  V^{-1}\!\!\!\int_{V(\rtilde)}\!\ddint{r'}
\left(
  \ln \left(
  1 + \sum_{\nvec} \mu_{\nvec}(\rvec') e^{\imag \kvec_{\nvec} \rvec'}
  \right) - 1 \right)
\rho_0 \left(
  1 + \sum_{\nvec} \mu_{\nvec}(\rvec') e^{\imag \kvec_{\nvec} \rvec'}
\right)
- \Phi_0  \\
&&+ \int_\Omega \ddint{r'} C_0 \rho_0
-\half \rho_0^2
\int_\Omega\ddint{\rtildeInt} \ V^{-1}\!\!\!\int_{V(\rtilde)}
\!\ddint{r'} \int_\Omega\ddint{r''}  C^{(2)}_0(\rvec''-\rvec')
\sum_{\nvec \mvec} \mu_\nvec(\rvec') \mu_\mvec(\rvec'')
e^{\imag (\kvec_{\nvec} \rvec' + \kvec_{\mvec} \rvec'')}
\nonumber
\end{eqnarray}
As in the calculation of the Fourier coefficients of
$C^{(2)}$, see \Eref{def_ck}, the equality between the single integral over
$\Omega$ and the double integral over individual volumes $V(\rtilde)$
centered at $\rtilde$ ignores surface terms and
holds only in the limit of infinite or periodic
$\Omega$, because the volume $V(\rtilde)$ at a point $\rtilde$ close to
the surface of $\Omega$ might not be fully within $\Omega$.
After expanding the coefficients $\mu_\mvec(\rvec'')$ about $\rvec'$ and
using the definition \Eref{def_ck} (in the notation $\nabla_k
c(-\kvec_\mvec)=\nabla_k|_{-\kvec_\mvec} c(\kvec)$ and noting that
$c(-\kvec)=c(\kvec)$ etc. ) the last triple integral becomes
\begin{equation}
V'' \int_\Omega\ddint{\rtildeInt}  V^{-1}\!\!\!\int_{V(\rtilde)}
\!\ddint{r'}
\sum_{\nvec \mvec} \mu_\nvec(\rvec') \left[
   \mu_\mvec(\rvec') c(\kvec_\mvec)
-  \imag (\nabla_\rvec \cdot \nabla_\kvec) \mu_\mvec(\rvec') c(\kvec_\mvec)
-  \half (\nabla_\rvec \cdot \nabla_\kvec)^2 \mu_\mvec(\rvec') c(\kvec_\mvec)
+ \ldots
\right]
e^{\imag (\kvec_{\nvec} + \kvec_{\mvec}) \rvec'}
\end{equation}
which is very similar to the right-hand side of
\Eref{self_consistency_expanded}. Assuming a perfect separation of
length scales, so that the $\mu_\nvec(\rvec')$ do not change within
the volume of a unit cell $V$, allows us to replace
$\mu_\nvec(\rvec')$ by $\mu_\nvec(\rtilde)$ within the integrals
over $V(\rtilde)$ and to make use of the orthogonality of the
exponentials, so that $\WpotNoneq$ becomes
\begin{eqnarray}
&& \WpotNoneq([\rho_i],[u_i\equiv\beta\mu]) \nonumber\\
&=&\int_\Omega\ddint{\rtildeInt} V^{-1}\!\!\!
\int_{V(\rtilde)}\!\ddint{r'} \left(
  \ln \left(
  1 + \sum_{\nvec} \mu_{\nvec}(\rtilde) e^{\imag \kvec_{\nvec} \rvec'}
  \right) - 1 \right)
\rho_0 \left(
  1 + \sum_{\nvec} \mu_{\nvec}(\rtilde) e^{\imag \kvec_{\nvec} \rvec'}
\right)
\elabel{WpotNoneq_final} \\
&&- \Phi_0  + \int_\Omega \ddint{r'} C_0 \rho_0 \nonumber \\
&&-\half \rho_0^2
V'' \int_\Omega\ddint{\rtildeInt}
\sum_{\nvec} \mu_{-\nvec}(\rtilde) \left[
   \mu_\nvec(\rtilde) c(\kvec_\nvec)
-  \imag (\nabla_\rvec \cdot \nabla_\kvec) \mu_\nvec(\rtilde) c(\kvec_\nvec)
-  \half (\nabla_\rvec \cdot \nabla_\kvec)^2 \mu_\nvec(\rtilde) c(\kvec_\nvec)
+ \ldots
\right]
\nonumber
\end{eqnarray}
This is the final result for the grand potential. Differentiating
with respect to $\mu_{-\jvec}(\rhat)$ gives
\begin{eqnarray} \elabel{derivative_WpotNoneq}
\frac{\delta}{\delta \mu_{-\jvec}(\rhat)} \WpotNoneq([\rho_i],[u_i\equiv\beta\mu]) &=&
V^{-1}\int_{V(\rhat)}\ddint{r'} \rho_0 e^{\imag \kvec_{-\jvec} \rvec'}
  \ln \left(
  1 + \sum_{\nvec} \mu_{\nvec}(\rhat) e^{\imag \kvec_{\nvec} \rvec'}
  \right) \\
&&-\rho_0^2
V''
\left[
\mu_{\jvec}(\rhat) c(\kvec_{\jvec})
- \imag (\nabla_\rvec \cdot \nabla_\kvec) \mu_{\jvec}(\rhat) c(\kvec_\jvec)
-  \half (\nabla_\rvec \cdot \nabla_\kvec)^2 \mu_\jvec(\rhat) c(\kvec_\jvec)
+ \ldots \right]
\nonumber
\end{eqnarray}
using, again, $c(\kvec_{\jvec})=c(\kvec_{-\jvec})$,
$\nabla_k c(\kvec_{\jvec})=-\nabla_k c(\kvec_{-\jvec})$ etc. and
$\kvec_{-\jvec}=-\kvec_{\jvec}$. Moreover,
we made use of
\begin{equation}
\frac{\delta}{\delta f(x)} \int\dint{y} f'(y) g(y)=-g'(x)
\end{equation}
which is obtained by ignoring surface term in an integration by
parts. This is particularly important when trying to include into a
functional the first derivatives, $\nabla_\rvec \mu_\nvec(\rvec)$,
reminiscent of a Rayleigh friction term in a mechanical
interpretation.

\Eref{derivative_WpotNoneq} is to be compared to
\Eref{self_consistency_family}, the original result by Haymet and
Oxtoby, which is reproduced by \Eref{derivative_WpotNoneq} by
requiring $\frac{\delta}{\delta \mu_{-\jvec}(\rhat)}
\WpotNoneq\equiv0$. The time evolution of $\mu_{\jvec}(\rhat)$ is
usually set equal to that derivative,
\begin{equation}
\dot{\mu}_{\jvec}(\rhat) = - M \frac{\delta}{\delta \mu_{\jvec}(\rhat)} \WpotNoneq
\end{equation}
with mobility $M$ (which can be absorbed into the definition of time),
driving the system to the above mentioned root. This is precisely the
mechanism used in many phase field models.

Further simplifications are needed to write the
grand potential \Eref{WpotNoneq_final} in  Allen-Cahn form. The set of
independent amplitudes $\mu_\nvec(\rtilde)$ is replaced by
a single scaling amplitude $\phi(\rtilde)$ by writing
\begin{equation}
\mu_\nvec(\rtilde) = \phi(\rtilde) \mu^0_\nvec
\elabel{def_phi}
\end{equation}
with constant amplitudes $\mu^0_\nvec$ chosen to represent a crystalline phase, so that $\phi(\rtilde)=1$
corresponds to the crystalline phase, while $\phi(\rtilde)=0$ suppresses
all structure, corresponding to a liquid with density $\rho_0$.  The
field $\phi$ therefore is to be interpreted as the order parameter,
proportional to the local amplitude of the atomic density waves, \ie
$\phi$ is the \emph{crystallinity}. The common amplitude $\phi(\rtilde)$ is necessarily real, because
$\rho$ being real implies $\mu_{\nvec}(\rvec)^*=\mu_{-\nvec}(\rvec)$
and together with $\mu_{\nvec}^{0*}=\mu_{-\nvec}^0$ (see below) we therefore have
$\phi(\rtilde)^*=\phi(\rtilde)$.

Collecting all local contributions in the density $w(\phi(\rvec))$, the
grand potential \Eref{WpotNoneq_final} simplifies to
\begin{eqnarray}
\WpotNoneq([\phi]) &=& \int_\Omega\ddint{r} w(\phi(\rvec))
\elabel{WAC_first_step} \\
&& +\half \rho_0^2 V'' \int_\Omega\ddint{\rtildeInt}
\sum_{\nvec}  \phi(\rtilde) \mu^0_{-\nvec}
   \left[
     \imag (\nabla_\rvec \cdot \nabla_\kvec)  \phi(\rtilde) \mu^0_\nvec c(\kvec_\nvec)
     + \half (\nabla_\rvec \cdot \nabla_\kvec)^2 \phi(\rtilde) \mu^0_\nvec c(\kvec_\nvec)
\right]
\nonumber
\end{eqnarray}
which can be simplified further by ignoring the surface terms of the
integral $\int \ddint{\rtildeInt} \phi(\rtilde) \nabla_\rvec
\phi(\rtilde)$ so that the remaining integral containing derivatives of
$\phi(\rvec)$ has the structure
\begin{equation}
\int_\Omega \ddint{\rtildeInt}
\phi(\rtilde)
\sum_{\nvec} \mu^0_{-\nvec}\mu^0_\nvec
(\nabla_\rvec \cdot \nabla_\kvec)^2
\phi(\rtilde)
c(\kvec_\nvec) \ .
\elabel{square_grad_term}
\end{equation}
To simplify this expression, the set of reciprocal lattice vectors
$\kvec_\nvec$ in the sum must be reduced. We use the index $\nvec^0$
to indicate that we are considering only a subset
$\{\kvec_{\nvec^0}\}$ of all possible reciprocal lattice vectors
$\{\kvec_{\nvec}\}$. In the simplest
approximation, the set is reduced to the set comprising
only the shortest (non-vanishing) reciprocal lattice vectors, $\{\kvec_{\nvec^0}\}$.
This set $\{\kvec_{\nvec^0}\}$ forms a ``star''
which means that elements of this set are related by point symmetry operations
of the reciprocal lattice (for further details see the Appendix, \Subsref{stars}), so that the star is invariant under these operations.
If the
real-space lattice is FCC, for example, the reciprocal lattice is BCC and
the eight nearest neighbor reciprocal lattice
vectors $1\,1\,1$, $\overline{1}\,1\,1$, \ldots,
$\overline{1}\,\overline{1}\,\overline{1}$ form the star of shortest length vectors.
By including only the shortest reciprocal
lattice vectors the symmetry and lattice constant of the crystalline structure are described
correctly by the truncated Fourier expansion.
The approximation can be systematically improved by including stars
of longer reciprocal lattice vectors.

Symmetry requires that all coefficients $\mu^0_{\nvec^0}$ as introduced in
\Eref{def_phi} associated with any of the vectors $\kvec_{\nvec^0}$ within the
same star have equal magnitude. Inversion symmetry in the real-space lattice
ensures that $\mu^0_{-\nvec^0}=\mu^0_{\nvec^0}$, if we impose that the center
of inversion coincides with a site.  Since
$\mu^{0*}_{\nvec^0}=\mu^0_{-\nvec^0}$ by reality of $\rho$, all
$\mu^0_{\nvec^0}$ are real and therefore equal, $\mu^0_{\nvec^0}=\mu_0$.
It is this equality which is needed to simplify \Eref{square_grad_term}, by
allowing us to place
$\mu^0_{-\nvec}\mu^0_\nvec$ in front of the remaining sum, which
is dealt with in the following.

The direct correlation function of the bulk liquid is isotropic,
$c(\kvec)=\tilde{c}(|\kvec|)$, so that the second derivative with
respect to wave-vector components $1\le\alpha\le d$ and $1\le\beta\le d$
in \Eref{square_grad_term}
can be written as
\begin{equation}
\partial_{k_\alpha}\partial_{k_\beta} c(\kvec) =
  \frac{\delta_{\alpha\beta}}{|\kvec|} \tilde{c}'(|\kvec|)
- \frac{k_\alpha k_\beta}{|\kvec|^3} \tilde{c}'(|\kvec|)
+ \frac{k_\alpha k_\beta}{|\kvec|^2} \tilde{c}''(|\kvec|) \ .
\end{equation}
In a sum of the form
\begin{multline}
\sum_{\kvec_{\nvec^0}} (\nabla_\rvec \cdot \nabla_\kvec)^2 \phi(\rtilde)
c(\kvec_{\nvec^0}) =
\sum_{\kvec_{\nvec^0}}
\sum_{\alpha=1}^d \sum_{\beta=1}^d
\partial_{k_\alpha}\partial_{k_\beta}
\partial_{\xtilde_\alpha}\partial_{\xtilde_\beta}
\phi(\rtilde)
c(\kvec_{\nvec^0}) \\
=
\sum_{\kvec_{\nvec^0}} \sum_{\alpha,\beta}^d \left(
\frac{\delta_{\alpha\beta}}{|\kvec_{\nvec^0}|}
  \tilde{c}'(|\kvec_{\nvec^0}|) -
\frac{k_{\nvec^0,\alpha}k_{\nvec^0,\beta}}{|\kvec_{\nvec^0}|^3}
  \tilde{c}'(|\kvec_{\nvec^0}|) +
\frac{k_{\nvec^0,\alpha}k_{\nvec^0,\beta}}{|\kvec_{\nvec^0}|^2}
  \tilde{c}''(|\kvec_{\nvec^0}|) \right)
\partial_{\xtilde_\alpha}\partial_{\xtilde_\beta}
\phi(\rtilde)
\elabel{compo_sum_simplification} \ ,
\end{multline}
where the $\kvec_{\nvec^0}$ run over all elements of the star
$\{\kvec_{\nvec^0}\}$, the contributions of off-diagonal elements,
$\alpha\ne\beta$, vanish after taking the summation over the star
$\kvec_{\nvec^0}$, if one assumes a cubic crystal system.  Only the diagonal
elements remain and by symmetry each Cartesian component contributes
$\sum_{\{\kvec_{\nvec^0}\}} k_\alpha^2 = (q/d) |\kvec|^2$, where
$q=|\{\kvec_{\nvec^0}\}|$ is the cardinality of the star, \ie the number of
elements in $\{\kvec_{\nvec^0}\}$.  The dimension $d$ is the dimension of the
space spanned by the star.  For example, the $8$ nearest neighbors in a BCC
lattice (the reciprocal lattice of an FCC lattice) would have $q=8$ and $d=3$.

The proof of this simplification is based on the Great Orthogonality Theorem of
group theory \cite{InuiTanabeOnodera:1996} as detailed in the Appendix. The assumption
of cubic symmetry ensures that there is always a three-dimensional, unitary,
irreducible representation of the point group. The
assumption can be lifted if one allows for anisotropic terms (see the
Appendix,
\Subsref{stars}),
reflecting the anisotropy of the lattice. In the following, we consider only
cubic crystal systems.

The sum in the integrand of \Eref{square_grad_term} now becomes
\begin{eqnarray}
\sum_{\kvec_{\nvec^0}} \phi(\rtilde) \mu^0_{-\nvec^0}
(\nabla_\rvec \cdot \nabla_\kvec)^2 \phi(\rtilde) \mu^0_{\nvec^0} c(\kvec_{\nvec^0})
& = &
\left(
  \frac{q (d-1)}{|\kvec_0| d} \tilde{c}' (|\kvec_0|)
+\frac{q}{d}                 \tilde{c}''(|\kvec_0|) \right) \mu_0^2
\phi(\rtilde) \nabla_\rvec^2 \phi(\rtilde) \elabel{simplification} \\
& = & -\epsilon_{\kvec_0} \phi(\rtilde) \nabla_\rvec^2 \phi(\rtilde)
\nonumber \ ,
\end{eqnarray}
where we have used $|\kvec_0|$ to denote the magnitude of any member of the star $\{\kvec_{\nvec^0}\}$, and we
have introduced the coupling $\epsilon_{\kvec_0}$ defined as
\begin{equation}
\epsilon_{\kvec_0}=-
\left(
  \frac{q (d-1)}{|\kvec_0| d} \tilde{c}' (|\kvec_0|)
+\frac{q}{d}                 \tilde{c}''(|\kvec_0|) \right) \mu_0^2 \ .
\elabel{eps_dpd}
\end{equation}
The expression for $\epsilon_{\kvec_0}$ simplifies further because the structure factor \Eref{structure_factor} usually
peaks at the shortest reciprocal lattice
vector\cite{RamakrishnanYussouff:1979}, so that $\tilde{c}'$ vanishes and
$\tilde{c}''<0$. In that case
\begin{equation}
\epsilon_{\kvec_0}=
 \frac{q}{d}                 | \tilde{c}''(|\kvec_0|) |\mu_0^2 \ .
\end{equation}
which enters the grand potential \Eref{WAC_first_step} as follows
\begin{equation}
\WpotNoneq_{\text{AC}}([\phi]) = \int_\Omega\ddint{r}
\left(
w(\phi(\rvec))
 + \quarter \rho_0^2 V''
\epsilon_{\kvec_0}
\left(\nabla_\rvec \phi(\rvec)\right)^2\ \right) \ ,
\elabel{WpotNoneq_solid_liquid}
\end{equation}
ignoring
surface contributions again.
Functional differentiation of \Eref{WpotNoneq_solid_liquid} with respect to the
phase field $\phi(\rvec)$ then reproduces the Allen-Cahn equation
\cite{BoettingerETAL:2002}
\begin{equation}
\dot{\phi} = - M \left(
\frac{d w}{d \phi} -
\half \rho_0^2 V''
\frac{g}{d} |\tilde{c}''(|\kvec_0|)| \mu_0^2
\nabla_\rvec^2 \phi(\rvec)
\right)
\end{equation}
in which the relaxational assumption
\[
\frac{\delta}{\delta \phi} \WpotNoneq = -M \frac{d}{dt}\phi \ ,
\]
(with mobility $M$)
has been made to drive $\WpotNoneq[\phi]$ to a minimum with respect to $\phi$.

The above analysis may be improved by including more stars of reciprocal lattice vectors than just the
shortest. Indeed, the entire reciprocal lattice can be decomposed into
disjoint stars without double-counting.
The sum over more than one star $\{\kvec_{\nvec^0}\}$ produces an effective coupling
\begin{equation}
\epsilon = \sum_{\{\kvec_{\nvec^0}\}} \epsilon_{\kvec_{\nvec^0}}
\end{equation}
replacing $\epsilon_{\kvec_0}$ in \Eref{WpotNoneq_solid_liquid}.

\subsection{Grain boundaries}
As noted in \Sref{App_to_Ints} we confine ourselves to misorientations
where there is a CSL with a relatively small three-dimensional unit cell
to ensure that the finite Fourier domain $V(\rvec)$ in the separation of
length scales (see
\Eref{def_scalar_product}) does not lead to significant errors. In
practice this limits the treatment to grain boundaries in cubic
lattices, where such CSLs arise frequently.

In the simplest treatment of a grain boundary we consider one set of
shortest length reciprocal lattice vectors in each crystal, which we
call $\Sset_l=\{\kvec^l_{\nvec^0}\}$ and $\Sset_r=\{\kvec^r_{\nvec^0}\}$ for the left
and right crystals respectively. Again, both these sets form stars, see
Appendix, \Subsref{stars}, and they are related by the rotation that generates the
misorientation in the bicrystal. The atomic densities deep in the left and right crystals
are described by density waves with wave vectors that are elements of
these stars, and with equal amplitudes $\mu^l_{\nvec^0}=\mu^l_0$ and
$\mu^r_{\nvec^0}=\mu^r_0$ respectively, similar to the situation in the
crystal-liquid interface.  That does not mean, however, that \emph{all}
wave-vectors have non-zero amplitudes: For example, if a wave-vector in
the right star is not also member of the left star, then its amplitude
$\mu^l_{\nvec^0}$ vanishes identically by symmetry. Amplitudes are
non-zero and equal within the respective stars.
Again, it is this equality that is used in the following to simplify the
expressions.

In the spirit of \Eref{def_phi}, we approximate the spatial dependence
of the amplitudes
$\{\mu_{\nvec^0}(\rtilde)\}$ by introducing a phase field $\phi(\rtilde)$ for the grain boundary in the form
\begin{equation}
\mu_{\nvec^0}(\rtilde) = \left(1-\phi(\rtilde)\right) \mu^l_{\nvec^0} +
\phi(\rtilde) \mu^r_{\nvec^0}\ .
\elabel{def_gb_phi}
\end{equation}
To satisfy the boundary conditions far from the boundary plane
$\phi(\rtilde)$ must vary from $0$ deep in the left-hand crystal to $1$
deep in the right-hand crystal. But this form of
$\{\mu_{\nvec^0}(\rtilde)\}$ does not allow for density waves which have
non-zero amplitudes only in the grain boundary core, thereby restricting
the degrees of freedom available to the system. There is no such
restriction in place at the starting point of the derivation,
\Eref{WpotNoneq_final}, which included, in principle, all
$\kvec$-vectors of the direct sum of the two stars, \ie the entire ``DSC
lattice'' \cite{SuttonBalluffi:1995}.

At closer inspection, the parameterization \eref{def_gb_phi} of
$\mu_{\nvec^0}(\rtilde)$ makes a slightly different use of the
``crystallinity'' $\phi(\rtilde)$, as it forces the system to be, on
average, a superposition of both crystalline lattices: As one is
gradually ``switched off'', the other is gradually ``switched on'', so
that $\phi(\rtilde)$ is the crystallinity of the right lattice and
$1-\phi(\rtilde)$ is the crystallinity of the left lattice. Reciprocal lattice
vectors common to both lattices are predicted by \Eref{def_gb_phi} to be
constant throughout the bicrystal, if $\mu_0^l=\mu_0^r$. Even if
$\mu_0^l\ne\mu_0^r$ the parameterization \eref{def_gb_phi} disallows the
possibility of a disordered grain boundary structure
where all the $\{\mu_{\nvec^0}(\rtilde)\}$ are locally zero. To resolve
this problem, an alternative form of \Eref{def_gb_phi} is needed, for
example $\mu_{\nvec^0} = (1/2) \phi
\left(
(1+\phi) \mu^l_{\nvec^0}
+
(1-\phi) \mu^r_{\nvec^0}
\right)
$
with $\phi$ varying from $-1$ to $1$. Yet, by continuity such a form would \emph{force} the
grain boundary to be disordered somewhere \cite{TangCarterCannon:2006}. In the following, we will use
\Eref{def_gb_phi}, which leads to the Allen-Cahn equation in a natural
way.

To see how the two different stars enter, the following derivation is
presented in some detail.
Inserting \Eref{def_gb_phi} in \Eref{WpotNoneq_final} produces the sum
\begin{equation}
\sum_{\kvec_{\nvec^0}}
\left[\left(1-\phi(\rtilde)\right) \mu^l_{\nvec^0} + \phi(\rtilde)\mu^r_{\nvec^0}\right]
\left( \nabla_\rvec \cdot \nabla_\kvec \right)^2
\left[\left(1-\phi(\rtilde)\right) \mu^l_{\nvec^0} + \phi(\rtilde)\mu^r_{\nvec^0}\right]
c(\kvec_\nvec^0)
\elabel{gb_simplication_step_one}
\end{equation}
which can be rewritten as four sums,
\newcommand{\bigsum}[1]{\sum_{#1}}
\begin{equation}
\begin{array}{rlrll}
 &\bigsum{\kvec_{\nvec^0}} & \left(1-\phi\right) \mu^l_{\nvec^0}
  & \left( \nabla_\rvec \cdot \nabla_\kvec \right)^2
  & \left(1-\phi\right) \mu^l_{\nvec^0} \\
+&\bigsum{\kvec_{\nvec^0}} & \left(1-\phi\right) \mu^l_{\nvec^0}
  & \left( \nabla_\rvec \cdot \nabla_\kvec \right)^2
  & \phi \mu^r_{\nvec^0} \\
+&\bigsum{\kvec_{\nvec^0}} & \phi \mu^r_{\nvec^0}
  & \left( \nabla_\rvec \cdot \nabla_\kvec \right)^2
  & \left(1-\phi\right) \mu^l_{\nvec^0} \\
+&\bigsum{\kvec_{\nvec^0}} & \phi \mu^r_{\nvec^0}
  & \left( \nabla_\rvec \cdot \nabla_\kvec \right)^2
  & \phi \mu^r_{\nvec^0}  \ .
  \elabel{gb_simplication_step_one_B}
\end{array}
\end{equation}
In the grand potential these sums are integrated over all space, see
\Eref{square_grad_term}.
A term of the form $\phi(\nabla_\rvec \cdot \nabla_\kvec)^2(1-\phi)$ therefore equals
$(1-\phi)(\nabla_\rvec \cdot \nabla_\kvec)^2\phi$ following integration by
parts and ignoring surface terms, which in turn equals $-\phi(\nabla_\rvec \cdot
\nabla_\kvec)^2\phi$ after ignoring surface terms again. The first sum
in \eref{gb_simplication_step_one_B}, for example, then reads
\begin{equation}
\begin{array}{rlrll}
&\bigsum{\kvec_{\nvec^0}} & \left(1-\phi\right) \mu^l_{\nvec^0}
  & \left( \nabla_\rvec \cdot \nabla_\kvec \right)^2
  & \left(1-\phi\right) \mu^l_{\nvec^0} \\
=&\bigsum{\kvec_{\nvec^0}} & \phi \mu^l_{\nvec^0}
  & \left( \nabla_\rvec \cdot \nabla_\kvec \right)^2
  & \phi \mu^l_{\nvec^0} \\
=&\bigsum{\Sset_l} & \phi \mu^l_{\nvec^0}
  & \left( \nabla_\rvec \cdot \nabla_\kvec \right)^2
  & \phi \mu^l_{\nvec^0}
\end{array}
\end{equation}
where the last equality is due to the coefficients $\mu^l_{\nvec^0}$
vanishing for those $\kvec$-vectors that are not member of the left
star. For those $\kvec$-vectors that are members, on the other hand, the
corresponding coefficients $\mu^l_{\nvec^0}$ are all equal to $\mu^l_0$,
which can be placed outside the sum. Applying the same procedure to all
four sums in \Eref{gb_simplication_step_one_B}, gives\footnote{It is
very instructive to repeat the derivation, using the identity
$\sum_{\Sset_l\cup\Sset_r}=\sum_{\Sset_l} + \sum_{\Sset_r} -
\sum_{\Iset}$, where the last term accounts for the double counting due
to the first two terms.}
\begin{eqnarray}
 && \sum_{\Sset_l\cup\Sset_r}
   \left[\left(1-\phi(\rtilde)\right) \mu^l_{\nvec^0} + \phi(\rtilde)\mu^r_{\nvec^0}\right]
   \left( \nabla_\rvec \cdot \nabla_\kvec \right)^2
   \left[\left(1-\phi(\rtilde)\right) \mu^l_{\nvec^0} + \phi(\rtilde)\mu^r_{\nvec^0}\right]
\elabel{gb_simplication_step_three} \\
&=&\ \mu_0^{l\,2} \sum_{\Sset_l} \phi(\rtilde) (\nabla_\rvec \cdot \nabla_\kvec)^2 \phi(\rtilde)
 + \mu_0^{r\,2} \sum_{\Sset_r} \phi(\rtilde) (\nabla_\rvec \cdot \nabla_\kvec)^2 \phi(\rtilde)
 -2 \mu_0^l\mu_0^r \sum_{\Iset}
       \phi(\rtilde) (\nabla_\rvec \cdot \nabla_\kvec)^2 \phi(\rtilde)
       \ . \nonumber
\end{eqnarray}
The intersection $\Iset=\Sset_l\cap\Sset_r$ contains the
reciprocal lattice vectors common to both stars, for which the product
$\mu^l_{\nvec^0}\mu^r_{\nvec^0}$ is non-zero. The intersection is assumed to be itself a
star (see the Appendix, \Subsref{stars} for details). To
\Eref{gb_simplication_step_three} the simplification
based on the Great Orthogonality Theorem (see the Appendix) can be
applied.
Because the space spanned by $\Iset$ is potentially only a sub-vector
vector space of $\Rset^d$, a projection matrix $P_\Iset$ needs to be
introduced, which projects any vector of $\Rset^d$ to this sub-vector
space. For the cubic systems we are focusing on here, this sub-vector
space has either dimension $d_\Iset=0$ in which case $P_\Iset=0$, or $d_\Iset=d$ in which
case $P_\Iset=\ID$ is the identity matrix, or $d_\Iset=1$ in which case one
spatial direction, say $z$, can be chosen to coincide with the common
direction. Using the Great Orthogonality Theorem,
\Eref{gb_simplication_step_three} becomes
\begin{align}
 & \sum_{\Sset_l\cup\Sset_r}
   \left[\left(1-\phi(\rtilde)\right) \mu^l_{\nvec^0} + \phi(\rtilde)\mu^r_{\nvec^0}\right]
   \left( \nabla_\rvec \cdot \nabla_\kvec \right)^2
   \left[\left(1-\phi(\rtilde)\right) \mu^l_{\nvec^0} +
   \phi(\rtilde)\mu^r_{\nvec^0}\right]c(\kvec_{\nvec^0}) \nonumber \\
=&\ (\mu_0^{l\,2} + \mu_0^{r\,2})
       \phi(\rtilde)
       \left(
       \frac{q (d-1)}{|\kvec_0| d} \tilde{c}' (|\kvec_0|)
      +\frac{q}{d}                 \tilde{c}''(|\kvec_0|)
       \right) \nabla_\rvec^2 \phi(\rtilde) \elabel{gb_simplication_step_four} \\
& - 2\mu_0^l \mu_0^r
       \phi(\rtilde)
       \left(
        \frac{q_\Iset}{|\kvec_0|  }        \tilde{c}' (|\kvec_0|)  \nabla_\rvec^2
       -\frac{q_\Iset}{|\kvec_0| d_\Iset}  \tilde{c}' (|\kvec_0|)  \nabla_\rvec P_\Iset \nabla_\rvec
       +\frac{q_\Iset}{d_\Iset}            \tilde{c}''(|\kvec_0|)  \nabla_\rvec P_\Iset \nabla_\rvec
       \right) \phi(\rtilde) \nonumber \ ,
\end{align}
where the stars $\Sset_l$ and $\Sset_r$ each contain $q$ elements,
$\Iset$ contains $q_\Iset$ elements, and $d_\Iset$ is the dimension of the
space spanned by $\Iset$. The two terms in
\Eref{gb_simplication_step_four} have the same prefactor if
$\mu_0^l=\mu_0^r=\mu_0$, which is a physically sensible choice we adopt
henceforth.

Suppose there is no misorientation between the crystal lattices.
Then there is no grain boundary and \Eref{gb_simplication_step_four}
should reduce to zero. With no misorientation we would have
$\Sset_l=\Sset_r=\Iset$, so that $q_\Iset=q$, $d_\Iset=d$ and
$P_\Iset=\ID$. Substitution of these values into the right hand side of \Eref{gb_simplication_step_four} shows that
it does indeed vanish.

We identify in \Eref{gb_simplication_step_four} an \emph{isotropic} coupling term in which
\begin{equation}
\epsilon_i
= - 2\mu_0^2
       \left(
       \frac{q (d-1)}{|\kvec_0| d} \tilde{c}' (|\kvec_0|)
      +\frac{q}{d}                 \tilde{c}''(|\kvec_0|)
       \right)
  + 2\mu_0^2
       \left(
       \frac{q_\Iset}{|\kvec_0|  }        \tilde{c}' (|\kvec_0|)
       \right) \elabel{eps_cup}
\end{equation}
multiplies $-\phi(\rtilde)\nabla_\rvec^2\phi(\rtilde)$. There is also
an possibly \emph{anisotropic} coupling term multiplying $-\phi(\rtilde)\nabla_\rvec P_\Iset \nabla_\rvec\phi(\rtilde)$:
\begin{equation}
\epsilon_a
= 2\mu_0^2
       \left(
       -\frac{q_\Iset}{|\kvec_0| d_\Iset}  \tilde{c}' (|\kvec_0|)
       +\frac{q_\Iset}{d_\Iset}            \tilde{c}''(|\kvec_0|)
       \right)  \elabel{eps_cap} \ .
\end{equation}
This last term is not necessarily anisotropic, because for the
particular choice of stars, $P_\Iset$ might be $0$ or the identity. In a
cubic system, if $P_\Iset$ is non-zero and non-trivial for a given pair
of stars,
it is equal to any non-zero, non-trivial $P_\Iset$ produced by any other
pair of stars. If the approximation is improved by adding further pairs
of stars, they will all generate the same types of terms, namely either
multiplying $\nabla_\rvec^2$
or $\nabla_\rvec P_\Iset \nabla_\rvec$.

The resulting grand potential has again square gradient form
\begin{equation}
\WpotNoneq_{\text{AC}}([\phi]) = \int_\Omega\ddint{r}
\left(
w(\phi(\rvec))
 + \quarter \rho_0^2 V''
\left(
\epsilon_i
\left(\nabla_\rvec \phi(\rvec)\right)^2
+
\epsilon_a
\nabla_\rvec \phi(\rvec) P_\Iset \nabla_\rvec \phi(\rvec)
\right)
\right) \ ,
\elabel{WpotNoneq_solid_solid}
\end{equation}
We note the difference in sign for the two couplings,
\Eref{eps_cup} and \Eref{eps_cap}, which
becomes most obvious when $\tilde{c}'=0$.
Thus, we have shown that \emph{common reciprocal lattice vectors reduce the square gradient term in the interfacial
energy, resulting in a lower interfacial energy and a smaller
interfacial width},
the characteristic scale of which is given by the coupling
$\epsilon_i+\epsilon_a$, which has
the dimension of a length squared. The smaller this coupling, the
smaller the penalty for rapid changes in $\phi$.
The amount of the reduction by common reciprocal lattice vectors
depends on the magnitude of the second derivative $\tilde{c}''(|\kvec_0|)$, which decreases with
increasing $|\kvec_0|$. Thus the shorter the common reciprocal lattice vectors the larger their influence on the
width and energy of the boundary. The influence of common reciprocal lattice vectors on the boundary energy
has been known for a long time \cite{SuttonBalluffi:1995}, but we are not aware that their influence on the
boundary width has been noted before.

\section{Discussion}
In this paper we have shown how the classical density functional theory
of Haymet and Oxtoby may be modified to produce an Allen-Cahn type free
energy functional first for crystal-liquid interfaces and then for grain
boundaries. For both types of interface the phase field is identified
with the amplitudes of atomic density waves, providing a physical
interpretation of the ``crystallinity'' of phenomenological phase field
models.

Let us consider first the approximations that were required to derive an Allen-Cahn
free energy functional from classical density functional theory.
To obtain \Eref{self_consistency_expanded} only two approximations were
made. Firstly, the excess free energy functional was expanded and
the resulting series was truncated at the second order term, see
\Eref{2ndorder}.
Secondly the Taylor
series for $\mu_\nvec(\rvec')$ about $\rvec$ in
\Eref{self_consistency_expanded} was also truncated at the second order term.
In principle both these expansions may be taken to higher order terms, although
not without a significant increase in complexity.
This would be
the natural way to extend a phase field model to higher order terms with
parameters related to fundamental quantities such as higher order correlation
functions.

The key step which lead to equation \eref{WpotNoneq_final}
for the grand potential $\WpotNoneq$, and eventually to a phase field model,
was the separation of length scales.
However, this was \emph{not} an approximation as we discussed after \Eref{self_consistency_family}.

The approximations mentioned so far were those of classical density
functional theory of crystal-liquid interfaces as implemented by Haymet and Oxtoby.
To obtain a phase field model for a crystal-liquid interface two further approximations were
necessary. Firstly, the spatial dependence of all $\mu_\nvec(\rvec)$ was assumed to be described by a single
field, the ``phase field'' $\phi(\rvec)$.  While classical density
functional theory naturally handles a larger set of density waves
with independent amplitudes $\mu_\nvec$, phase field modeling in
single-component systems to
date has considered only a single order parameter, which we have
identified as being proportional to the
local amplitude of the atomic density waves.
The obvious deficiency is the highly restricted nature of the configurational phase space made
available to the crystal-liquid interface. When the same approach is adopted for a grain boundary,
see \Eref{def_gb_phi}, the restrictions are even more severe. 
Secondly, all $\mu_{\nvec^0}^l$ and $\mu_{\nvec^0}^r$ within a star are
assumed to be identical and real, which suppresses any relative phase
factor of the density waves.
For example, if there were a rigid
translation of one crystal relative to the other on either side of the grain boundary this would
lead to complex density wave amplitudes in one crystal. The simplistic assumption made in
\Eref{gb_simplication_step_three} eliminates the possibility of any rigid body relative translation.

It is clear that there are quite severe limitations of current
phase field models of grain boundaries in single-component systems,
at least when they are interpreted in the framework of classical
density functional. Can we use DFT to indicate how these phase field
models may be improved? We have already seen that going beyond the
second order term in the expansion \Eref{2ndorder} will introduce
higher order correlation functions and hence more information about
structure and bonding. Indeed, going to at least third order
correlation functions would seem to be necessary in covalent crystals
such as silicon. This would introduce higher order terms in the
grand potential \Eref{WpotNoneq_final} and in the Allen-Cahn free
energy functional \Eref{WpotNoneq_solid_liquid} with parameters that are directly
related to the higher order correlation functions. But perhaps the
most obvious and pressing improvement would be the introduction of a
second independent phase field to describe the amplitudes of two
sets of density waves, one for each crystal. Far from the interface
the crystal structures are related by a rotation. If we also wish to
be able to predict a relative rigid body translation between the
crystals, with components parallel and perpendicular to the
interface, it will be necessary to allow the density wave amplitudes
of one crystal to be complex.

\section*{Acknowledgements}
APS gratefully acknowledges the support of a Royal Society Wolfson
Merit Award. GP gratefully acknowledges the support of a Research Councils UK Fellowship. This
work has been supported by the European Commission under Contract
No. NMP3-CT-2005-013862 (INCEMS).

\appendix*

\section{Summing over a star}
\slabel{GOT}
In this Appendix, we derive the equations based on the Great
Orthogonality Theorem, as used in \Eref{simplification} and
\eref{gb_simplication_step_four}. In particular, we
will identify their prerequisites and possible extensions.

The aim is to simplify expressions of the form
\begin{equation}
\sum_{\alpha \beta} \sum_{\kvec_{\nvec^0}}
  k_{\nvec^0,\alpha} k_{\nvec^0,\beta}
  v_\alpha w_\beta
= \sum_{\alpha \beta} M_{\alpha \beta}
  v_\alpha w_\beta
\elabel{simplify_compo}
\end{equation}
where we may interpret $M_{\alpha \beta}$ as an element of a matrix $M$. Introducing
bra and ket notation for convenience the matrix $M$ can be expressed as,
\begin{equation}
\sum_{\kvec_{\nvec^0}} \ket{\kvec_{\nvec^0}} \bra{\kvec_{\nvec^0}} = M
\ .
\elabel{simplify_matrix}
\end{equation}
In Eqns.~\eref{simplify_compo} and \eref{simplify_matrix} the sums run
over a set of vectors $\{\kvec_{\nvec^0}\}\subset\Rset^d$ of equal
length, which, for the time being, we will call a ``star''. This term
will be defined properly in the next section.

The star is picked from the reciprocal lattice, and it is expected to
display some point group symmetries of this lattice, in the sense that these
point group operations effect only a permutation of members of the star, so that the star itself
remains invariant.  We assume that the star under consideration is
invariant under a group of order $g$, which has a unitary, irreducible representation
$\{U_n\}$, $n=1,\dots,g$, and that this representation has the same
dimension $d$ as the vectors that make up the star.  Acting with $U_n$
from the left and with $U_n^\dagger$ from the right on both sides of
\Eref{simplify_matrix} merely amounts to a permutation of the summands,
because of the invariance of the star. Therefore $M$ 
commutes with any group element,
\begin{equation}
U_n M U_n^\dagger=M
\ ,
\elabel{invar_M}
\end{equation}
so that $\sum_n U_n M U_n^\dagger=g M$ and therefore
\begin{equation}
g M_{kl} =
\sum_n^g \sum_{i,j}^d
(U_n)_{k i}
M_{ij}
(U_n^\dagger)_{jl}
=
(\Tr M) \frac{g}{d} \delta_{kl}
\end{equation}
where in the last equality we have used the Great Orthogonality Theorem
(for unitary representations),
\begin{equation}
\sum_n^g (U_n)_{k i} (U_n^\dagger)_{jl} = \frac{g}{d} \delta_{k l}
\delta_{i j} \ .
\end{equation}
By construction, the trace $\Tr M$ is the sum over the squares of the
moduli of all vectors in the star, which all have the same length, which we call
$|\kvec_0|$. So, if the cardinality of the star is $q$, then $\Tr M = q
|\kvec_0|^2$ and one finally arrives at the general result
\begin{equation}
\sum_{\kvec_{\nvec^0}} k_{\nvec^0,\alpha} k_{\nvec^0,\beta}
= \frac{q}{d} |\kvec_0|^2 \delta_{\alpha \beta}
\elabel{simplication_by_GOT_V1}
\end{equation}
provided there exists a group under which the star is
invariant and for which there is a unitary, irreducible representation of the same
dimension as the vectors that make up the star.

The fact that the Great Orthogonality Theorem applies only to \emph{irreducible}
representations of the group is
a limitation of the above derivation, which motivates the
following extension. By construction, the
rank of the matrix $M$ is $d$, which is the dimension of the vector space the
$\kvec_{\nvec^0}$ are taken from. \emph{A priori}, it is unknown whether
there exists a group under which the star is invariant and that has an
\emph{irreducible}, unitary representation with dimension $d$.
\Eref{simplication_by_GOT_V1} applies only if a suitable group and a
suitable representation exists.

If the star spans a sub vector space of dimension $d'$ less than $d$,
one cannot expect that it is invariant under a representation with
dimension $d$. For example, the star $\Kset_1=\{1\,0\,0, 0\,1\,0,
\overline{1}\,0\,0, 0\,\overline{1}\,0\}$ is \emph{not} invariant
under \emph{any} irreducible three-dimensional representation of
\emph{any} group, because it will always contain some group elements
that will rotate the star out of the $xy$-plane in which all members of the star are located. So,
\Eref{simplication_by_GOT_V1} cannot apply to $\Kset_1$.

On the other hand, considering only the  two-dimensional sub vector
space spanned by $\Kset_1$ (the $xy$ plane), the star $\Kset_2=\{1\,0, 0\,1, \overline{1}\,0,
0\,\overline{1}\}$ derived from $\Kset_1$ by appropriate projection, is
invariant under some unitary, irreducible representations of $C_{4v}$, some of
which have dimension $2$ (and some have dimension $1$). Of course, the
original star $\Kset_1$ is also invariant under some three-dimensional
representations of $C_{4v}$, but none of them is irreducible.

If a star is contained entirely in a vector space of dimension $d'<d$, its members can be
expressed in $d$-dimensional space as
\begin{equation}
\ket{\kvec_{\nvec^0}} = O B \ket{\kvec'_{\nvec^0}}
\qquad\text{and}\qquad
\bra{\kvec_{\nvec^0}} = \bra{\kvec'_{\nvec^0}} B^\dagger O^\dagger
\elabel{k_from_kprime}
\end{equation}
where $O$ is an arbitrary rotation matrix of rank $d$, $B$ is a $d\times d'$
projection matrix
with $B_{ij}=\delta_{ij}$ and
$\ket{\kvec'_{\nvec^0}}$ is a vector in $\Rset^{d'}$. The matrix $B$
increases the number of elements in the vector from $d'$ to $d$ and the
rotation matrix $O$ rotates the resulting vector to an arbitrary
position.
The vectors
$\{\ket{\kvec'_{\nvec^0}}\}$ also form a star, now in the sub-vector space
$\Rset^{d'}$, so that
\begin{equation}
\sum_{\kvec'_{\nvec^0}} \ket{\kvec'_{\nvec^0}} \bra{\kvec'_{\nvec^0}}
= \frac{q}{d'} |\kvec'_0|^2 \ID
\elabel{braket_prime}
\end{equation}
if there exists a group under which the star $\{\ket{\kvec'_{\nvec^0}}\}$ is
invariant, and which has a (unitary) irreducible representation of dimension
$d'$.
Therefore
\begin{equation}
\sum_{\kvec_{\nvec^0}} \ket{\kvec_{\nvec^0}} \bra{\kvec_{\nvec^0}}
= \frac{q}{d'} |\kvec_0|^2 (OBB^\dagger O^\dagger)
\elabel{general_res_explicit}
\end{equation}
using $|\kvec_0|=|\kvec'_0|$ and applying $OB$ on \Eref{braket_prime}
from the left and $B^\dagger O^\dagger$ from the right. The matrix
$(OBB^\dagger O^\dagger)=P$ (rotate, remove element, rotate
back) projects a vector in $R^d$ onto the
vector space isomorphic to $R^{d'}$.
For example
\begin{equation}
P=\left(
\begin{array}{rrr}
1 & 0 & 0 \\
0 & 1 & 0 \\
0 & 0 & 0
\end{array}
\right)
\end{equation}
for the star $\Kset_1$ introduced above, using its representation in the
$xy$ plane, $\Kset_2$. The construction of the vectors $\kvec_{\nvec^0}$
from $\kvec'_{\nvec^0}$ in \Eref{k_from_kprime} represents an important
limitation: At first sight, \Eref{general_res_explicit} seems to apply to every star
that has some degree of symmetry. However, not every star can
be written in the form \Eref{k_from_kprime}. For example,
$\Kset_3=\{1\,0\,1, 0\,1\,1,
\overline{1}\,0\,1, 0\,\overline{1}\,1\}$ has a $C_{4v}$ symmetry, but this
star spans the entire $\Rset^d$ and so there is no star
$\{\ket{\kvec'_{\nvec^0}}\}\subset\Rset^{d'}$ that would fulfill
\Eref{k_from_kprime} with $d'<d$.

In a more compact form, the general result is
\begin{equation}
\sum_{\kvec_{\nvec^0}} k_{\nvec^0,\alpha} k_{\nvec^0,\beta}
= \frac{q}{d'} |\kvec_0|^2 P_{\alpha \beta} \ ,
\elabel{simplication_by_GOT_V2}
\end{equation}
provided there exists an unitary, irreducible representation of a group under which the star
$\{\ket{\kvec_{\nvec^0}}\}\subset\Rset^d$ is invariant, with a
dimension $d'$ identical to that of the space spanned by the star.
Furthermore, if $d'=d$, then $P$ reduces to the identity.
\Eref{simplication_by_GOT_V2} can be used in the form
\begin{equation}
\sum_{\alpha \beta}
\sum_{\kvec_{\nvec^0}}
k_{\nvec^0,\alpha} k_{\nvec^0,\beta}
\partial_{\xtilde_\alpha} \partial_{\xtilde_\beta}
=
\frac{q}{d'} |\kvec_0|^2
\sum_{\alpha \beta}
P_{\alpha \beta}
\partial_{\xtilde_\alpha} \partial_{\xtilde_\beta}
=
\frac{q}{d'} |\kvec_0|^2 \nabla_{\xtilde} P \nabla_{\xtilde}
\end{equation}
which leads to \Eref{gb_simplication_step_four}.

\subsection{Definition of a star}
\slabel{stars}
So far in this Appendix a star has been defined as any set of $k$-vectors of equal length.
We now adopt the standard definition of a star, which will enforce the
conditions required for
\Eref{simplication_by_GOT_V1} or \eref{simplication_by_GOT_V2} to apply:
A star is generated from an
initial vector by applying all symmetry operations in the form of a
unitary, irreducible representation of the point group
of the lattice in the appropriate coordinate
system.\footnote{We do not allow for translational equivalence, \ie we
do not discount elements from a star which are generated by the
translation group. \cite{Tinkham:1964}} This representation is chosen so that the lattice
itself is invariant under its action, so that the star represents a
finite subset of the lattice. The star generated is highly degenerate,
giving rise to the so-called ``little groups''.

There are two important caveats: Firstly, only cubic crystals have point
group symmetries with three-dimensional irreducible representations.
Secondly, the intersection of two stars does not necessarily have a
irreducible with a dimension equal to that of the space spanned by the
intersection.  Generally, both points can be addressed in the same way,
namely by decomposing stars or the intersection thereof into smaller
stars which then have the required properties. But that produces
additional projection matrices, so that, for example, the isotropic
$\nabla_{\rtilde}^2$ terms in \Eref{simplification} and
\eref{gb_simplication_step_four} split into multiple terms with
different projections of the form $\nabla_{\rtilde} P \nabla_{\rtilde}$.

All the cubic point groups, and only the cubic point groups, have
three-dimensional, unitary, irreducible representations, and the
intersection of any two stars generated by one of these representations is either empty,
one-dimensional, or it coincides with both stars. For cubic crystals all equations
derived above apply without restrictions, producing an isotropic term
and a anisotropic one with a single, preferred direction.

\bibliography{articles,books}

\end{document}